\documentclass[10pt, a4paper]{article}
\usepackage{authblk}

\usepackage{cmap}
\usepackage[T2A]{fontenc}		
\usepackage[utf8]{inputenc}
\usepackage[english]{babel}
\usepackage{amsmath}
\usepackage{amssymb}
\usepackage{graphicx}
\usepackage{xcolor}
\usepackage{tabularray}
\UseTblrLibrary{booktabs}%%%
\usepackage{hyperref}
\usepackage[normalem]{ulem}
\usepackage{authblk}

\usepackage[left=2cm,right=2cm,
top=2cm,bottom=2cm,bindingoffset=0cm]{geometry}
\title{ \bf Multimodal reconstruction of TbCo thin film structure with Basyeian analysis of polarised neutron reflectivity}
\author{P.S.~Savchenkov$^{1,2}$,
        K.V.~Nikolaev$^{1,3}$,
        V.I.~Bodnarchuk$^{1,4,5}$,
        A.N.~Pirogov$^{6}$,
        \\
        A.V.~Belushkin$^{1,4,7}$
        and
        S.N.~Yakunin}

        \affil[1]{National Research Center Kurchatov Institute, Moscow, Russia}
        \affil[2]{National Research Nuclear University <<MEPhI>>, Moscow, Russia}
        \affil[3]{Moscow Institute of Physics and Technology, Dolgoprudny, Russia}
        \affil[4]{Frank Laboratory of Neutron Physics, Joint Institute for Nuclear Research, Dubna,
Russia}
        \affil[5]{Dubna State University, Dubna, Russia}
        \affil[6]{M.N. Miheev Institute of Metal Physics, Ural Branch of the Russian Academy of Sciences, Ekaterinburg, Russia}
        \affil[7]{ Kazan (Volga Region) Federal University, Insitute of Physics, Kazan, Russia}

\newcommand{\pvec}[1]{\vec{#1}\mkern2mu\vphantom{#1}}

\begin{document}
	
	\maketitle
	
	\begin{abstract}
            We implemented the Bayesian analysis to the polarised neutron reflectivity data.
            Reflectivity data from a magnetic TbCo thin film structure was studied using the bundle of a Monte-Carlo Markov-chain algorithm,
            likelihood estimation,
            and error modeling.
            By utilizing the Bayesian analysis,
            we were able to investigate the uniqueness of the solution beyond reconstructing the magnetic and structure parameters.
            This approach has demonstrated its expedience as several probable reconstructions were found (the multimodality case) concerning the isotopic composition of the surface cover layer.
            Such multimodal reconstruction emphasizes the importance of rigorous data analysis instead of the direct data fitting approach,
            especially in the case of poor statistically conditioned data,
            typical for neutron reflectivity experiments.
            The analysis details and the discussion on multimodality are in this article.
	\end{abstract}

\section{Introduction}

Neutron reflectivity (NR) is a unique tool employed in the study of surfaces and layered nanostructures.
By measuring specularly reflected low energy (thermal or slow) neutrons at grazing angles, one effectively probes the depth profile of the scattering length density (SLD) of the sample.
It is sensitive to both the nuclear and the magnetic SLDs.
Therefore, it is primarily advantageous compared to X-ray reflectivity experiments to the study to the study of magnetic properties of matter or the structure of matter with low Z contrast.
Thus, NR has a wide range of applications in science and technology.
In spintronics, it is used for metallic thin film characterization~\cite{zabel09, zhong20}.
In biophysics, NR is used to study biomembranes~\cite{majkrzak00,fragneto01}
and macromolecules~\cite{belivcka15} as deposited on the surface.
In chemical research, NR is employed to the study of polymers structure and surfactants ~\cite{braun17}.
NR has one of its most prominent applications in the study of atmospheric phenomena: the study of aerosols~\cite{jones16} and the study of oxidation kinetics of organic monolayers~\cite{sebastiani15}.
Additional information  of the spin-states of reflected neutrons allows  the reconstruction of the spatial  distribution and the orientation of magnetic moments. By using polarized neutron reflectometry (PNR) this method provides a more comprehensive understanding of the magnetic properties of materials~\cite{majkrzak91,ankner99}.

The reflectivity calculation is irreversible, i.e., 
one cannot directly extract SLD profiles from the NR data.
That is due to the loss of phase and the dynamic scattering effects like the total external reflection.
Thus, despite our understanding of NR physics, the reconstruction of the SLD profile from experimental data is a complicated problem.
Typically, SLD reconstruction from NR data is done by building a forward model to simulate reflectivity.
Further, instead of directly studying the data, one studies the behavior of the model as compared to the data.
Then, iteratively, using optimization algorithms, one varies the SLD profile until the corresponding reflectivity calculation fits the data according to the priorly chosen goodness of fit criterion.
Usually, it is the $\chi^2$ criterion(see, for example~\cite{jang2016accurate}).
The SLD corresponding to the optimal value of $\chi^2$ is the sought sample reconstruction.

There are two main obstacles to such reconstruction in the context of PNR.
First, PNR experiments yield data with a low signal-to-noise ratio compared to, for instance, X-ray reflectivity data.
That is due to a lower incident beam flux.
The use of the neutron reflectometry method requires high collimation and simultaneously imposes limitations on the cross-sectional area of the neutron beam. 
This significantly affects the neutron flux in the incident beam. 
Flux density values of 10$^5$ - 10$^6$ cm$^{-2}$ s$^{-1}$ are typical for many neutron reflectometers, while synchrotron sources typically have flux densities in the range of 10$^{13}$ - 10$^{14}$ cm$^{-2}$ s$^{-1}$. The low neutron flux leads to limited statistical significance of the measurement results. Alternatively, long beam times at the source are required to perform a single experiment. However, it is not always feasible to increase the experiment time for the sake of collecting statistics.
Still, the sensitivity of PNR to the nuclear composition and magnetic structure necessitates its use.
Therefore, the approach optimized for the low signal-to-noise data is required.
Second, the uniqueness of reconstruction is not guaranteed.
Moreover, in some cases, it is provably not unique~\cite{majkrzak1995exact}
(see also examples for the X-ray reflectivity in~\cite{zimmermann2000phase, zimmermann2006}).
Thus, one PNR dataset can be explained by several different SLD profiles -- the so-called multimodality.
Meanwhile, the optimization approach provides a single solution,
and its uncertainty is estimated locally, i.e., 
with no regard to other possible solutions.
That requires a comprehensive approach capable of identifying multiple probable SLD reconstructions.

We address both these issues by implementing the Monte-Carlo Markov chain (MCMC) algorithm
in the analysis of the PNR data.
This approach provides a numeric estimation of the likelihood of a given model, 
which is the probability that a particular model SLD corresponds to the observed PNR data.
Unlike the optimization approach, which yields only one optimal solution, 
MCMC yields a likelihood distribution for different SLD models across a specified range of parameters.
If there are several possible solutions, the likelihood will have several optima, 
and MCMC allows their investigation simultaneously.
As such, we were able to study PNR data for the multimodality.
Together with the MCMC approach, we implemented the error modeling similar to what described in~\cite{heidenreich2015bayesian}.
In some instances, there exists an optimal (most probable) measurement error correction, which MCMC can also study as part of the model.
In the case of a low signal-to-noise ratio, which is precisely the case with PNR data, the estimation of model parameters depends on the correct estimation of the measurement errors.

In this work, we focus on the PNR data analysis strategy rather than on the description of PNR as a physical phenomenon, as it is well studied; see an overview in \cite{ruhm99,blundell92} among others. On the contrary, the implementation of MCMC for the PNR data analysis is scarce in the literature, 
although it is getting more and more attention in the scientific 
community~\cite{sivia1991bayesian, sivia98, durant2021fisherinfo, mccluskey2023advice}.  
We implemented MCMC for the reconstruction of the SLD profile of a Tb$_{0.03}$Co$_{0.97}$ thin film structure from the PNR data.
The reconstruction revealed two possible solutions, 
emphasizing the importance of studying the data for multimodality,
particularly in statistically ambiguous data such as PNR.
\section{Deterministic and stochastic analysis of PNR data}

    There are several theoretical formalisms for calculating PNR\cite{ruhm99,blundell92}. 
    All of them are based on solving time-independent Schrodinger equation and matching  wave function and its derivative at the boundaries of thin films. In this study, we use matrix formalism\cite{blundell92} for the forward simulation ($I^{\rm calc}(\vec p)$ -- forward model).
    This model provides a simulation of reflected beam intensity for a thin film model described with a vector of parameters $\vec p$. 
 
    In this study we use a stochastic approach to analyze the PNR data.
    However, let us first revisit a conventional deterministic approach.
    The deterministic approach consists of three parts: a forward model,
    a goodness-of-fit criterion,
    and an optimization algorithm.
    Primarily, Pearson's $\chi^2$ is used (see examples in \cite{yakunin2014combined,andrle2021shape}) as a goodness-of-fit criterion:
    \begin{equation}
        \label{eq:chi2}
	\chi^2=
        \frac{1}{N-l}
        \sum_{i}
        \frac{\left[ I^{\rm calc}_i(\vec p)-I_i^{\rm exp}\right]^2}{\sigma^2_i},
    \end{equation}
    where $I^{\rm exp}$ is the measured reflectivity data,
    $N$ is the size of $I^{\rm exp}$ i.e. the number of measured points,
    $l$ is the number of free parameters used for the forward model,
    $\sigma$ -- statistical errors of experimental points.
    The sum in Eq.~\ref{eq:chi2} is taken over the data points.
    
    The $\chi^2$ statistics imply the sum of stochastic variables with the standard normal distribution: $\mathcal{N}(0,1)$. 
    Hence, by using $\chi^2$,
    one assumes that experimental data is normally distributed around some $I^{\rm calc}(\pvec{p}')$ with the standard deviation $\sigma$,
    such so $ [I^{\rm exp} - I^{\rm calc}(\pvec{p}')]/\sigma \sim \mathcal{N}(0,1)$.
    Here, $\pvec{p}'$ is the optimal parameter vector.
    In other words, 
    $\pvec{p}'$ represents the set of model parameters which best  describes the experimental data.
    Thus, finding $\pvec{p}'$ means reconstructing the structure parameters.
    It is evident from Eq.~\ref{eq:chi2} that 
    for $\pvec{p}'$ with increasing $N$
    the value converges $\chi^2 \to 1$.
    Indeed, if one assumes that the averaging over the realizations of a stochastic variable is equivalent to the averaging over the angular data points, then the Eq.~\ref{eq:chi2}
    is merely producing the variance of a standard normal distribution in the limit, which is 1.
    Thus, by minimizing the $\chi^2$ one finds a reconstruction $\pvec{p}'$ of the sample model. 
    Such a search is typically done via an optimization algorithm, 
    see e.g. ~\cite{dennis1981adaptive,powell2009bobyqa,more1978lm_algorithm}.

    The reconstruction uncertainty can be estimated using the Hessian $\mathbf H$ of $(N-l)\chi^2$ calculated at $\pvec{p}'$.
    Assuming the Eq.~\ref{eq:chi2} the elements of the Hessian can be approximated as
    \begin{equation}
        H_{jk} =
        (N-l)
        \dfrac{\partial^2\chi^2}{\partial p_j \partial p_k}
        \approx 
        \sum_i 
        \dfrac{2}{\sigma^2_i}
        \dfrac{\partial I_i^{\rm calc}}{\partial p_j}
        \dfrac{\partial I_i^{\rm calc}}{\partial p_k}
        .
    \end{equation}
    Here, the contribution of the higher-order derivative is negligible as it is multiplied by $(I^{\rm calc}-I^{\rm exp})$, 
    which is zero on average. 
    Then, again, assuming that the measurement errors are distributed normally,
    one can calculate the covariance matrix as the inverse of Hessian 
    $\mathbf C = \mathbf{H}^{-1}$.
    Finally, with the covariance matrix $\mathbf C$,
    one can estimate the uncertainty of reconstructed parameters.
    Now the standard deviation of parameter reconstruction is calculated from the diagonal elements: ${\rm STD}(p_j) = \sqrt{C_{jj}}$.
    Together with the reconstruction uncertainty, 
    covariance matrix $\mathbf{C}$ instantly gives the Pearson correlation matrix $\mathbf{R}$, 
    with elements $R_{jk} = C_{jk}/\sqrt{C_{jj}C_{kk}}$.
    This matrix describes the correlation between the reconstructed parameters,
    i.e., how much the  change in one parameter can be offset by the  change in another, 
    maintaining a similar goodness-of-fit. 
    For example, $R_{jk}>0$ means that a positive (or negative) shift in $p_j$
    can be compensated by a positive
    (or, correspondingly, negative)
    shift in $p_k$.
    Thereby, 
    $R$ provides the information on how well the optima $\pvec{p}'$ is defined. 
    It can also be considered from a geometry perspective. Consider a multi-dimensional space of parameters in which a point describes a specific model $\vec p$. Then, the optimal model $\pvec{p}'$ is a single point in this parametric space. Furthermore, the uncertainty is represented by a geometric $l$-dimensional shape surrounding $\pvec{p}'$ in this space. The optimization of $\chi^2$ yields an estimation of the optimum $\pvec{p}'$, the standard deviation of parameters yields the estimate of the size of an optimum in each dimension, and the correlation coefficient provides an estimate of how similar the shape of a 2D section of an optimum in parametric space to the line. If, for instance, $R_{ij} = \pm 1$, then the section of an optimum in the $ij$ plane is flat, and the model is over-defined. Therefore, one aims to find a model with lower correlations. In that sense, the result of such an analysis is represented by a 0D cross-section of a parametric space -- the solution itself, a 1D cross-section -- the error of reconstruction, and an estimate of a 2D cross-section -- correlation in model parameters. 
      
    One of the drawbacks of this approach is the assumption of the uniqueness of the solution, according to which the optimal set of parameters $\pvec{p}'$ minimizing the function $\chi^2 $ is considered to be the complete and only description of the reconstructed sample structure. At the same time, this approach fails to take into account the possibility of the influence of underestimation of measurement errors on the restored sample structure. This problem becomes particularly relevant in cases of limited statistical information or significant background noise in the experimental data. In this context, parameter sets corresponding to significantly different structures may be statistically close, so a more optimal $\chi^2$ value may characterize the non-physical solution. In this framework, it is impossible to establish the solution's uniqueness and correctly evaluate its stability. The application of the presented approach may need to be corrected in the case of the distribution of model parameters being different from the normal and nonlinear correlations between them. 

    %TODO: add stitching paragraph
  
    Another approach for solving the optimization problem used in this paper is a stochastic analysis, namely the Bayesian inference. This method is particularly effective with many-parameter models and a low signal-to-noise ratio. The approach was originally successfully applied to astrophysics and cosmology problems~\cite{dunkley05}.
    However, in recent years, its application has been extended to other areas of physics, such as X-ray and neutron reflectometry~\cite{mccluskey2023advice}.
    The details of the method can be found in literature \cite{heidenreich2015bayesian}.
    Let us briefly focus on its fundamental principles.
	 
    The posterior function (posterior probability density) $\pi(I|D)$ of the expected parameters of the structure $\vec p$, given already existing knowledge about the system $I \equiv I^{\rm calc}(\vec{p})$ and given the occurrence of experimental events $D$, by the Bayes' theorem, is:
    \begin{equation}
        \label{eq:bay}
        \pi(I|D)
        =
        \frac{\pi(D|I)\pi(I)}{\pi(D)}.
    \end{equation}
    The prior function ({\it apriori} probability density)
    $\pi(I)$ corresponds to the already available data on the system under the study.
    The denominator of the Eq.~\ref{eq:bay} $\pi(D)$
    describes the probability of observing experimental data $D$,
    thus in our case $\pi(D) = 1$.
    
    The function $\pi(D|I)$ determines the probability that the model calculated for the parameter set $\vec p$ agrees with the experimental data set $D$.
    Simply put, $\pi(D|I)$ is a measure of how much evidence there is to justify the use of a particular model.
    This function is called the likelihood $\mathcal{L} = \pi(D|I)$. 
    Assuming that the experimental data are measured with normally distributed errors,
    the logarithm of the likelihood function can be written in the form:
    \begin{equation}
        \label{eq:likelihood}
        \log\mathcal{L}
        =-\frac{1}{2}\sum_{i}
        \left\{
            \frac{\left[ I^{\rm calc}_i(\vec{p})-I_i^{\rm exp}\right]^2}{\sigma^2_i}+\log{2\pi \sigma^2}
        \right\}
        .
    \end{equation}
    The log-likelihood in its form and its underlying meaning coincides with the Eq.~\ref{eq:chi2} with the difference that $\log{\mathcal{L}}$ is not normalized by the number of experimental points is strictly negative, and has an additional term with the logarithm of measurement errors.
    The best agreement between the model and experimental now data corresponds to the maximum of the likelihood function as opposed to the minimum of $\chi^2$.  	  
 
    Using the stochastic approach to solve the optimization problem makes it possible to conduct a more thorough evaluation of experimental errors. Indeed, in addition to systematic and statistical errors, actual measurements are subject to inevitable measurement noise. There are two primary sources of noise in reflectometry measurements. First, variations in the intensity of the incident neutron beam, and second, noise from background fluctuations.
    Note that Eq.~\ref{eq:likelihood} has an optimum with respect to $\sigma$.
    In statistical terms, this means that there is a maximum plausible error. The error estimate can be corrected within the same structure reconstruction process if there is a correct systematic model for an error.
    Analysis in~\cite{heidenreich2015bayesian,karamanis2021zeus} showed that the linear error model is applicable for experiments with discrete signal detection:
    \begin{equation}
        \label{eq:errmod}
        \sigma^2=[a I^{\rm calc}(\vec{p})]^2+b^2+\sigma_e^2
        ,
    \end{equation}
    where $\sigma_e$ is the standard deviation, as in the case of the $\chi^2$ method, a and b are free scalar parameters included in the vector $\vec p$. Thus, in the presented approach, the error model is optimized together with the experiment model.
 	  
    One approach to approximating the posterior for a set of parameters is the Monte Carlo numerical modeling methods for Markov chains (MCMC) \cite{foreman13,karamanis21}. This approach allows to numerically obtain a set of parameters p that is statistically equivalent to the likelihood $\mathcal L$, 
    while it becomes possible to explore large parameter spaces efficiently \cite{sivia98}. One can estimate the parameter reconstruction error by analyzing the sample of each parameter from the vector p. By evaluating the pairwise realizations ($p_i$, $p_j$ ), one can numerically estimate the correlation between the $i$-th and $j$-th parameters without being limited by the assumption of its linearity. In this way, the correlation function of the parameters can be analyzed directly, as well as the assumption of the linear nature of the correlation. Another advantage of this approach is the absence of the requirement of the uniqueness of the solution, which makes it possible to consider multimodal posterior distributions (however, there remains the need to choose the most appropriate solution from several variants).
\section{Experiment layout  and sample growth }

\begin{figure}
	\centering
	\includegraphics[width=0.6\linewidth]{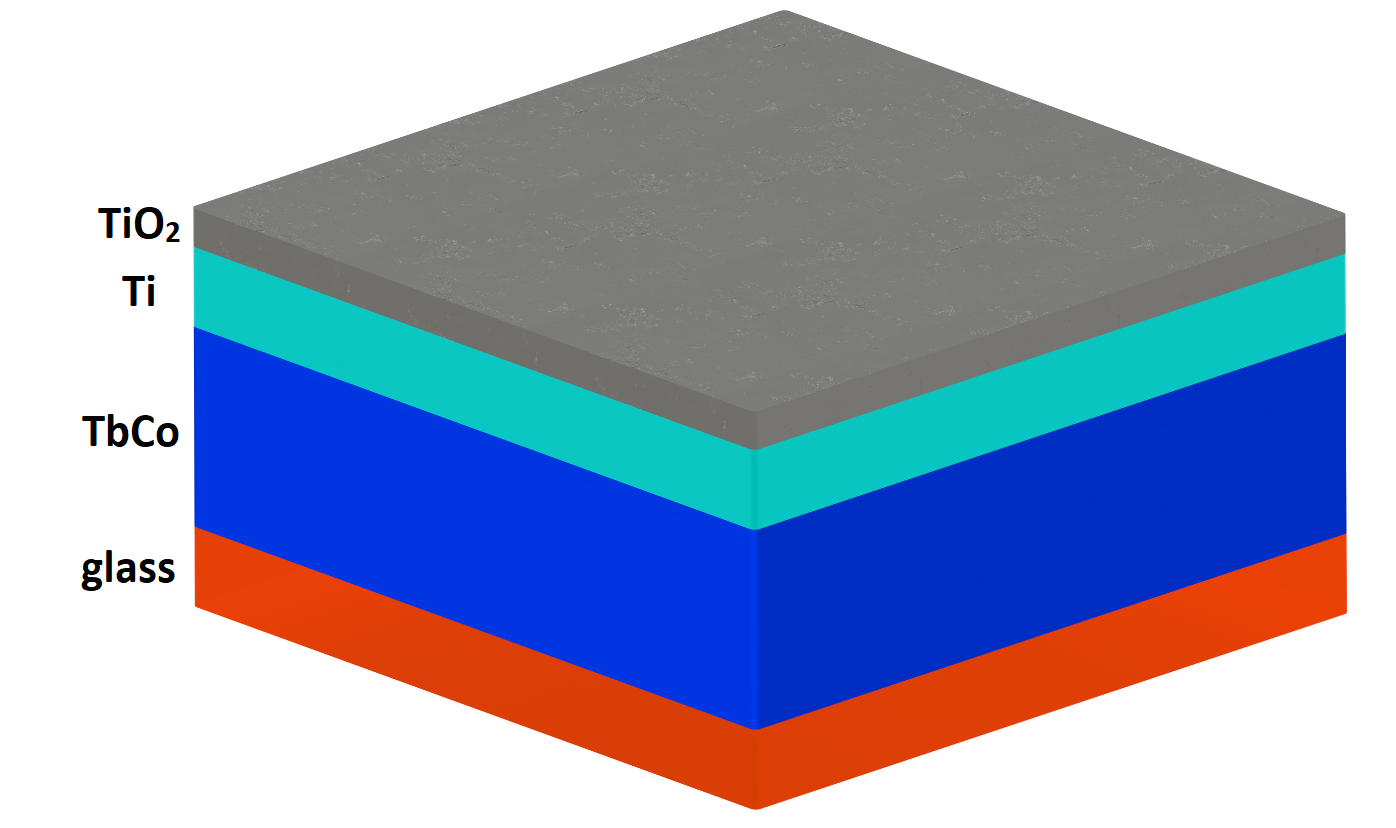}
	\caption{
		Nominal structure model. A Tb$_{0.03}$Co$_{0.97}$ layer with a nominal thickness of 1000\AA ~ is positioned on a SiO$_2$ substrate and is protected from atmospheric exposure by a partially oxidized Ti layer with a nominal thickness of 500 \AA. }
	\label{fig:struc}
\end{figure}

{\it Sample growth.}
The thin amorphous film Tb$_{0.03}$Co$_{0.97}$ was grown using magnetron sputtering machine. The nominal thickness of the Tb$_{0.03}$Co$_{0.97}$ layer is 1000 \AA. The sputtering gas was Ar of 99.9999\% purity and the growth pressure 10$^{-3}$ mbar. <<Corning>> glass substrate with size 20$\times$20 mm  was used. Uniform magnetic field of 150 Oe during growth, were applied parallel to the film plane to imprint a small uniaxial in-plane magnetic anisotropy. The sample had a protective coating of Ti thin layer (nominal thickness is 500 \AA), which was applied at the final stage of the process cycle. Due to the well-known oxidation of Titanium in the atmosphere, it is reasonable to assume the presence of an external TiO$_2$ layer. Thus, the model structure of the investigated thin film is shown in Figure \ref{fig:struc} and has the form: TiO$_2$/Ti/Tb$_{0.03}$Co$_{0.97}$/glass. 

{\it PNR data acquisition.}
Experimental measurements were carried out at the reflectometer REFLEX \cite{aksenov92} which is located at beamline 9 of the high flux pulsed reactor IBR-2. The reflectometer operates in time-of-flight mode with an available wavelength range of 1.2 - 8 \AA. The V-cavity polarizer (Swiss Neutronics AG)  provides an average polarization level of 90\% over the entire range. The momentum transfer range is 0.001 - 0.13 \AA$^{-1}$, and the sample flux is 10$^5$  cm$^{-2}$s$^{-1}$. The detector system consists of a two-dimensional position sensitive detector (PSD) with a size of 200$\times$200 mm$^{2}$ and a spatial resolution of 1.5 mm \cite{belushkin06,churakov18}. 

The feature of the beamline 9 is that it is a tangential channel with respect to the moderator, i.e. the radiating surface is the flat end part of the thermal moderator, whose main working surface is a neutron source for the 4th, 5th and 6th beamline. 
As a result, the REFLEX reflectometer is an instrument with a small flux of thermal neutrons on the sample. 

The experiment performed in full polarisation analysis mode requires three successive reflections of the neutron beam: from the polariser mirror, from the sample and from the analyser mirror. Each reflection leads to some broadening of the neutron beam due to scattering on the roughness and non-ideal flatness of the mirror surfaces. In the experiment performed, this fact, together with the small size of the sample, resulted in a rather blurred image of the scattered intensity on the position sensitive detector, slightly above the background level. In order to collect a sufficient number of counts for analysis within a reasonable amount of experimental time, it was necessary to increase the angular range within which data was collected. As a result, the resolution of the momentum transfer was degraded and it was not possible to determine its exact value during the experiment. However, it is possible to estimate the resolution from the setup parameters. Considering the neutron flash duration (Full Width at Half Maximum (FWHM) $\sim$  320~$\mu\text{s}$) and the distance from the moderator to the detector (42~m), the time-of-flight resolution is calculated to be about $\frac{3}{\lambda}\%$. Furthermore, the beam was collimated to about 0.3~mrad at a measurement angle of 4 mrad, which corresponds to an angular divergence of $\sim 8\%$. Consequently, the resolution of $\delta Q/Q$ is estimated to be in the range of 8-9\%.

\begin{figure}
	\centering
	\includegraphics[width=1.0\linewidth]{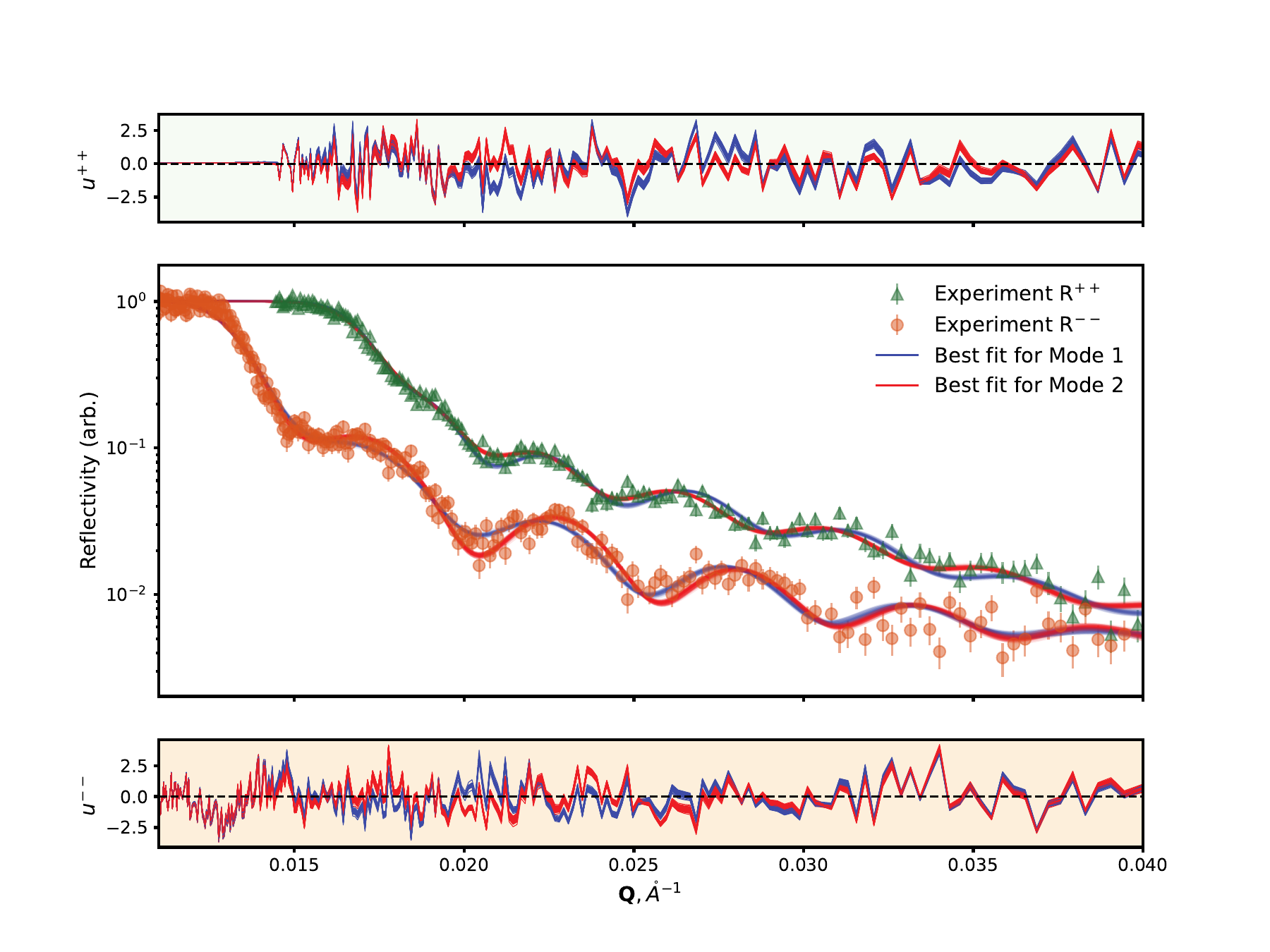}
	\caption{Experimental (points) and simulated (solid lines) polarised neutron reflectometry
		curves from Tb$_{0.03}$Co$_{0.97}$ sample after normalisation. Lines represent 500 samples from the posterior distributions. The bottom and top panels show residuals u$^{++}$ and u$^{--}$ for R$^{++}$ and R$^{--}$ respectively. }
	\label{fig:ref1}
\end{figure}

The result of the polarized neutron reflectometry experiments on the Tb$_{0.03}$Co$_{0.97}$ thin film is shown in Figure \ref{fig:ref1}. Here, the green triangles  and orange dots  correspond to the non-spin-flip reflectivities R$^{++}$ and R$^{--}$, respectively. The experimental data presented are normalised so that the reflectivity saturates at unity at low wave vector transfer (normalised by the incident neutron intensity).

\section{Results and discussion}

\begin{figure}
    \centering
    \includegraphics[width=0.8\linewidth]{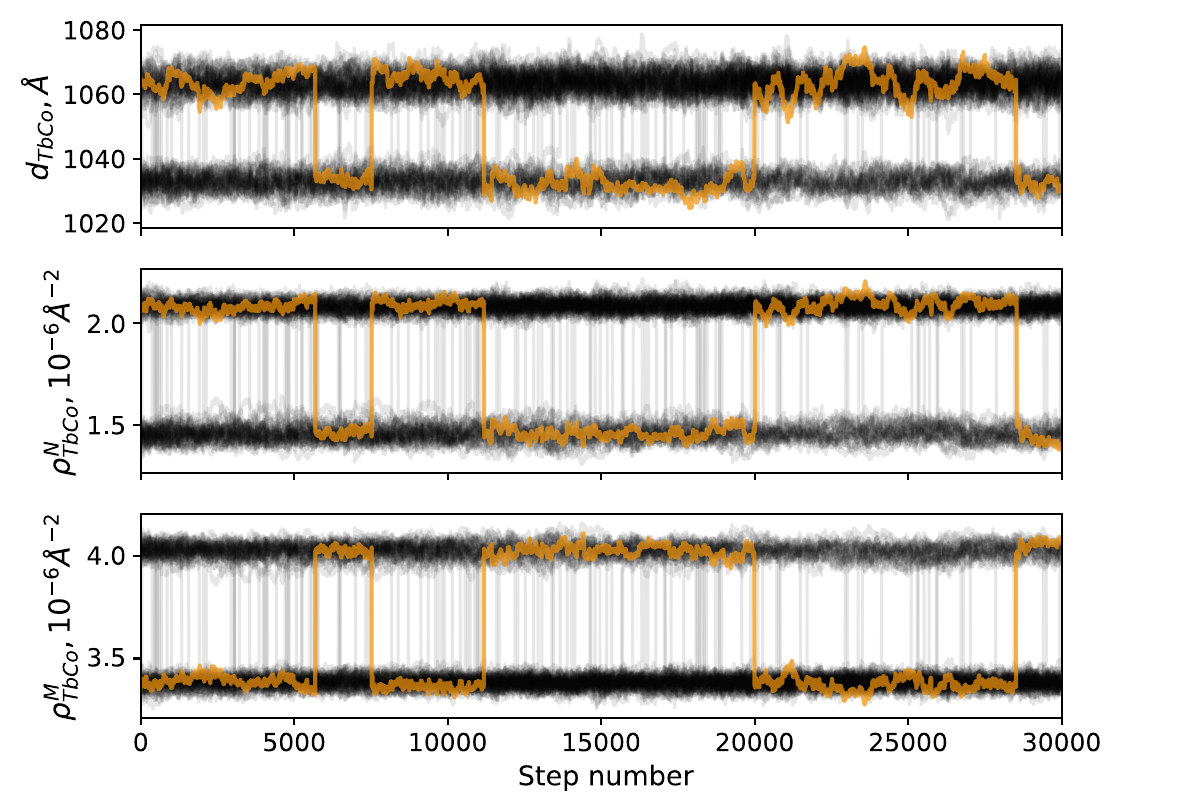}
    \caption{  The individual panels show the trace of the 30000 walker positions for the structural parameters of the Tb$_{0.03}$Co$_{0.97}$ layer. One randomly chosen walker is highlighted in yellow to show that the walkers are able to move around the parameter space and jump between two well separated modes. }
    \label{fig:walk}
\end{figure}

The thin film structure was reconstructed using an MCMC algorithm~\cite{foreman13} by analysing the log-likelihood function Eq.~\ref{eq:likelihood}. 
For the forward model, the structure was divided into two interface layers (labeled "In1" and "In2") at the surface of the structure, the magnetic layer (labeled "TbCo") and the glass substrate layer (labeled "SiO$_2$").
Each layer is parameterized by the thickness $d$, the nuclear contribution to the SLD $\rho^{\rm N}$.
Imperfections at the interlayer boundaries, which may result from sample growth peculiarities or interdiffusion, were considered as follows: A transition layer of width $\eta$ was assumed to exist between layers $a$ and $b$, within which the SLD varied according to the law: 
 \begin{equation}
        \label{eq:erff}
        \rho(z)=\rho_a+\frac{\rho_b-\rho_a}{2}
        \left[ 
            1+{\rm erf}
            \left(
                \frac{z-z_t}{\sqrt{2}\eta}
            \right)
        \right], 
    \end{equation}
where ${\rm erf}(z)$ is the error function, $z_t=(z_b-z_a)/2$ the position of the transition layer. 

The magnetic layer is also parameterized by its magnetic contribution to the SLD $\rho^{\rm M}$. A linear error model was considered to fit the estimation of the measurement uncertainties. As in Eq.~\ref{eq:errmod}, the parameters for the error model are $a$ and $b$. Finally, the instrumental resolution $\delta Q/Q$ and background noise values were also optimized.  
A total of 96 walkers were initialised, each with randomly generated starting values for layer thicknesses, SLD values and for transition layer thicknesses. The prior information used in the calculations included nuclear SLD $\rho^{\rm N}$ values for each layer in the range $[-6.0,6.0]\times$10$^{-6}$\AA$^{-2}$, the thicknesses $d$ and transition layer thicknesses $\eta$ of the two outermost layers limited to [0,200] \AA, and the thickness of the magnetic layer $d_{\rm TbCo}$ limited to $[800,1400]$~\AA. Optimisation parameters included background noise values within $[10^{-6}-10^{-2}]$ and resolution $\delta Q/Q$ within $[0.05,0.15]$. 

After setting the walkers, the burn-in phase is initiated. In this work, the model is burnt in for 5,000 steps to allow the walkers to distribute in the parameter space according to the posterior probability. Based on visual inspection of the walker trajectory, it was determined that this number of steps was sufficient for the walkers to converge. After a burn-in period of 5,000 steps, the production models were run for 30,000 steps. The chains were checked to ensure convergence by the estimate of the integrated autocorrelation time. 

Figure \ref{fig:walk} shows the trace of 96 walkers for the structural parameters of the Tb$_{0.03}$Co$_{0.97}$ layer over the 30,000 MCMC steps after the burn-in phase. It is obvious that the walkers have separated into two different positions. One randomly chosen walker is highlighted in yellow to show that the walkers are able to move around the parameter space and frequently jump from mode to mode. The behavior of the walkers indicates multimodality: there are two sets of model parameters for which the values of the likelihood function (\ref{eq:likelihood}) are comparable; during the calculations it was found that this difference does not exceed 2\%.

\begin{figure}
	\centering
	\includegraphics[width=1\linewidth]{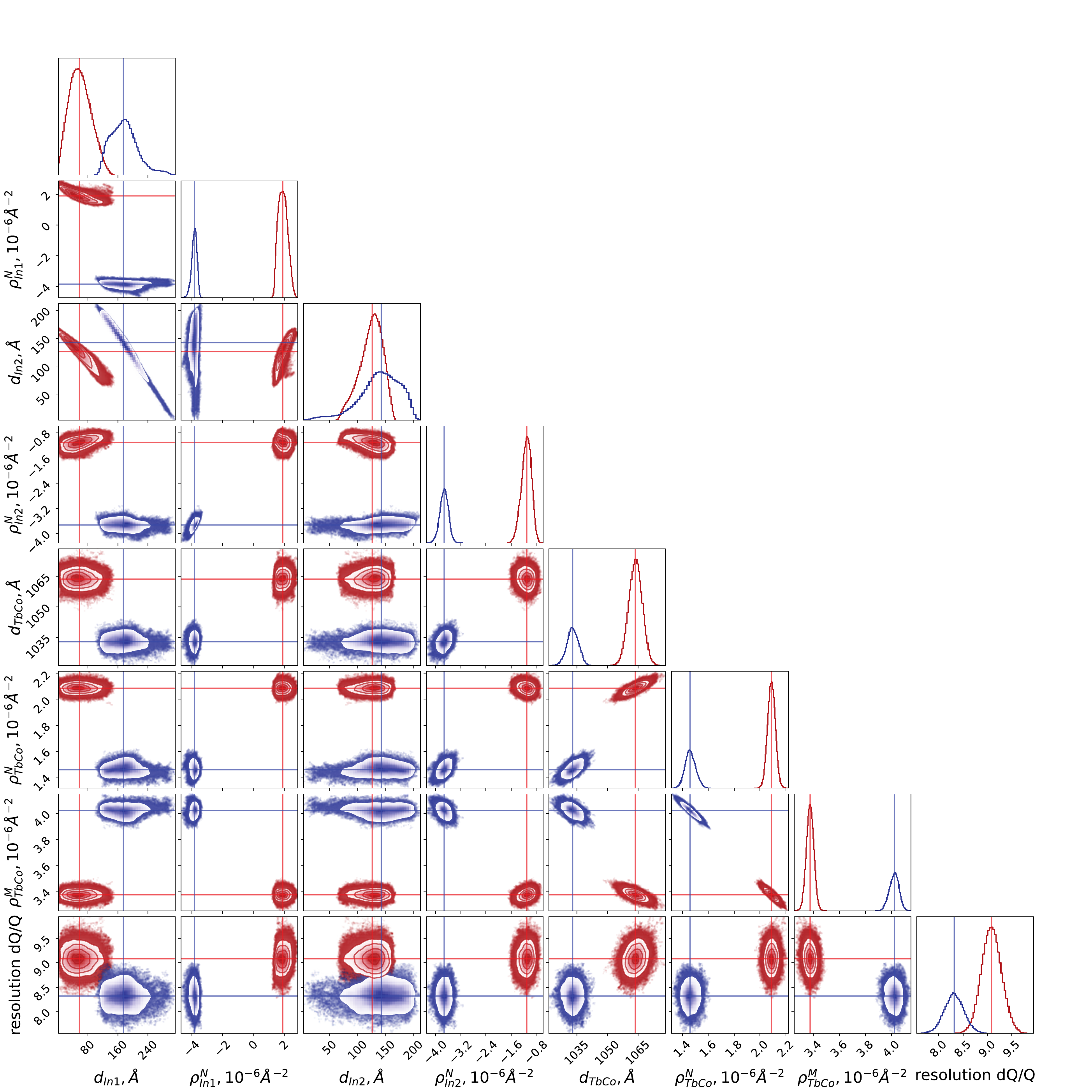}
	\caption{A corner plot of the MCMC sample. Two different modes are marked with colors: Mode 1 -- blue, Mode 2 -- red. Diagonal subplots are the 1D histograms. Other subplots are the pair plots showing the correlations between each pair of parameters. Visualization is done with the corner package~\cite{foreman2016corner}.}
	\label{fig:corner_plot}
\end{figure}

\begin{figure}
	\centering
	\includegraphics[width=0.7\linewidth]{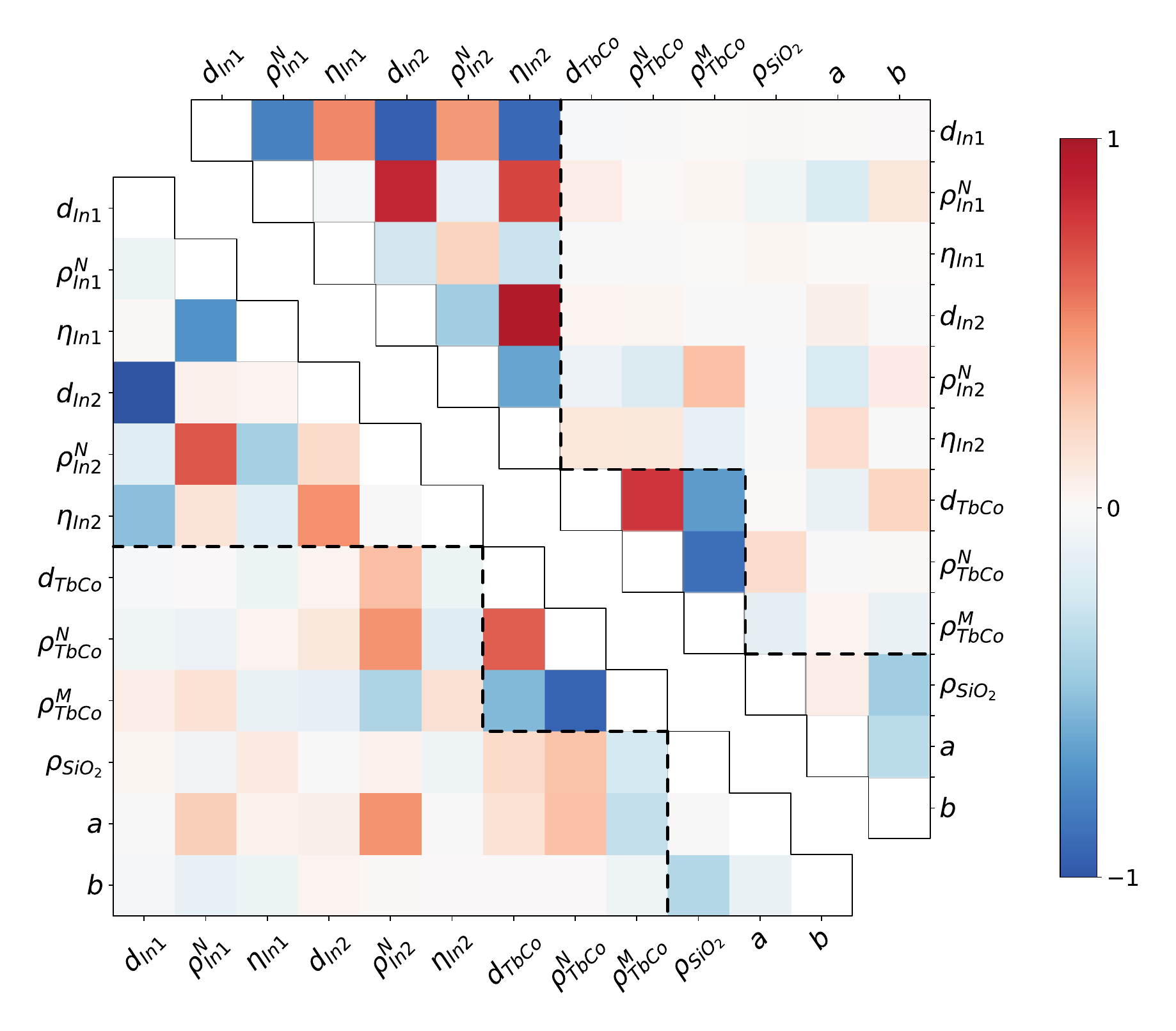}
	\caption{Matrix of Pearson's correlation coefficients between variables for Mode 1 (lower left corner) and for Mode 2 (upper right corner).}
	\label{fig:pearson}
\end{figure}

A corner plot of the posterior distributions for a selected set of the structural parameters and the instrumental resolution is shown in Fig.~\ref{fig:corner_plot}. 
The set of parameters selected for demonstration is not the full set that was optimized, as this figure would be difficult to read. For the full set of structural parameters, see Table~1, and for their correlations, see the Pearson correlation matrix in Fig.~\ref{fig:pearson} later in the text.
The diagonal panels show the 1D histogram for each model parameter, with a solid line indicating the median. The off-diagonal panels show 2D projections of the posterior probability distributions for each pair of parameters. It can be seen that the posterior distribution of all parameters has a two-peak structure. This indicates the presence of two modes of parameters with comparable values of the likelihood function. 
To make it easier, different solution modes is shown with different colors for better understanding. We will call the solution mode with a negative value of SLD $\rho^{\rm N}_{\rm In1}$  as Mode 1, represented by the blue color, and the one with a positive value of SLD $\rho^{\rm N}_{\rm In1}$ as Mode 2, represented by the red color. Based on the information provided in Figure \ref{fig:corner_plot}, it is clear that the posterior distribution deviates from normal for certain parameters. In such cases, the median (highlighted by lines in Figure \ref{fig:corner_plot}) becomes a more informative measure of central tendency than the mean.
The final result of the reconstruction are shown in Table~1.
We chose the median (50\% quantile) as the estimate of the structure parameter. For the reconstruction uncertainty interval, 16\% and 84\% quantiles are chosen.

\begin{table}
	\centering
        \makebox[0pt]{
	\begin{tblr}{
			hlines, vlines,
			colspec = {X[1.5,c] X[1.0,c] X[1.0,c]  X[1.0,c] X[1.0,c]  X[1.0,c] X[1.0,c]  X[1.0,c] X[1.0,c]},
			row{1} = {font=\bfseries},
                row{2} = {font=\bfseries},
		}
		\toprule
		Layer &  \SetCell[c=2]{c} {Thickness \\ $d$,~{\rm \AA}} & &\SetCell[c=2]{c}{Transition layer \\ thickness $\eta$, {\rm \AA}} & 
            &\SetCell[c=2]{c}{Nuclear SLD \\ $\rho^{N}$, {\rm 10$^{-6}$\AA$^{-2}$} } & & \SetCell[c=2]{c}{Magnetic SLD \\{\rm $\rho^{\rm M}$, 10$^{-6}$ \AA$^{-2}$} }  \\
		& {Mode 1} & {Mode 2} & {Mode 1} & {Mode 2}  & {Mode 1} &         {Mode 2}  & {Mode 1} & {Mode 2}  \\
		
  \midrule
		Interface 1 & 179$^{+44}_{-36}$ & 61$^{+34}_{-28}$  & 120$^{+6}_{-6}$ &   25$^{+26}_{-16}$   & -3.8$^{+0.2}_{-0.2}$ & 1.9$^{+0.3}_{-0.3}$ & -- & --\\
		Interface 2 & 137$^{+34}_{-46}$ & 124$^{+18}_{-25}$  &  47$^{+50}_{-32}$& 97$^{+28}_{-42}$  & -3.7$^{+0.2}_{-0.2}$ & -1.1$^{+0.1}_{-0.2}$  & -- & --\\
		Tb$_{0.03}$Co$_{0.97}$        & 1033$^{+3}_{-3}$ & 1063$^{+3}_{-3}$  &  104$^{+5}_{-5}$&  102$^{+4}_{-4}$ & 1.45$^{+0.04}_{-0.03}$ & 2.09$^{+0.02}_{-0.02}$  & 4.03$^{+0.03}_{-0.03}$ & 3.38$^{+0.01}_{-0.01}$\\
		SiO$_2$     & -- &  -- & \{50\} & \{50\}  &  3.32$^{0.01}_{-0.02}$&  3.38$^{+0.01}_{-0.01}$  & -- & --\\
		\bottomrule
	\end{tblr}
        }
	\caption{Magnetic and structural  parameters reconstructed from the PNR data using the MCMC.
                The results are given for both modes of the multimodal solution.
                Subscripts and superscripts indicate 16\% and 84\% quantiles respectively.
                Curly braces indicate fixed parameters.
                }
\end{table}

The results of the calculation using the transfer matrix method \cite{blundell92} of polarized neutron reflectometry curves for 500 samples from the posterior distributions  are shown in Figure \ref{fig:ref1} as solid lines.  The blue lines correspond to samples from the first solution mode, while the red lines correspond to the second. Furthermore, the residuals $u=(I^{\rm exp}-I^{\rm calc})/\sigma$ are also shown in Figure \ref{fig:ref1}.
The good agreement between the fit calculations and the experimental data can be seen for both solution modes from the residuals, which are well within the range of [-3,3].

\begin{figure}
	\centering
	\includegraphics[width=0.99\linewidth]{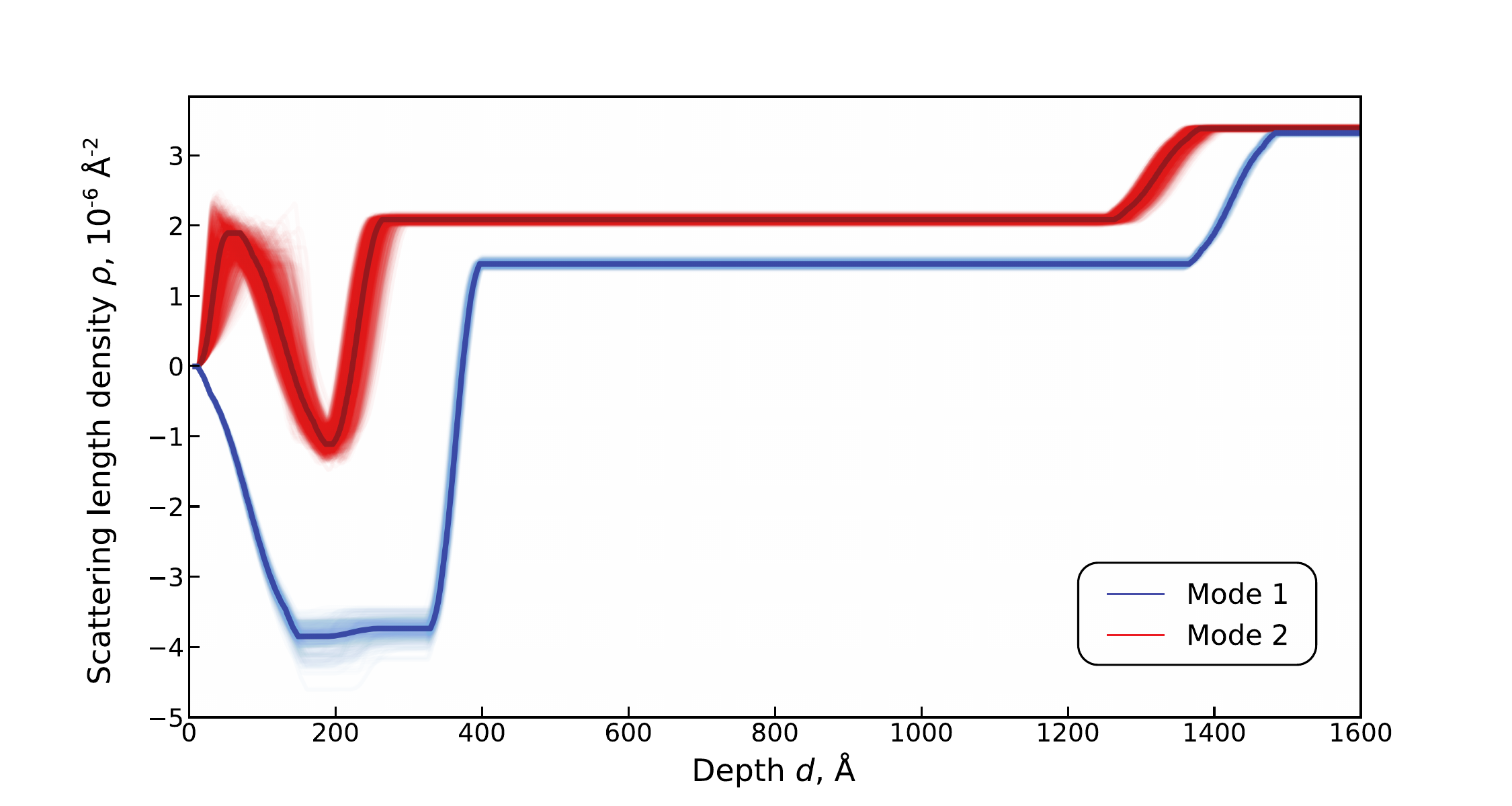}
	\caption{Reconstructed SLD profile of the Tb$_{0.03}$Co$_{0.97}$ film. The thin lines correspond to the values of the profile obtained from 500 randomly selected samples from the posterior distribution. The bold line represents the profile constructed using the median values.}
	\label{fig:profs}
\end{figure}

It is evident that the  SLD profiles reconstructed for each mode of the solutions show significant differences.  This is most evident in Figure \ref{fig:profs}, where the profiles for 500 randomly selected samples from the posterior distribution are represented by thin lines. In addition, the figure shows the profiles plotted for the median values of the parameters for each mode of the solutions.  
The primary difference between the SLD profiles obtained for different solution modes is observed in the outer  layers (see Table 1).

We would like to highlight an interesting result of this research.
The SLD values for the interfaces were sought in the wide range $[-6,\,6]\times 10^{-6}$~\AA$^{-2}$. 
Such choice of SLD during reconstruction was made without regard to the possible isotopic configuration of the interface material.
Nevertheless, the reconstructed interface SLDs in both Mode~1 and Mode~2 correspond to the tabular parameters of real materials, for the natural Ti $\rho^{\rm N} = -0.42$,
and for the $^{48}$Ti  $\rho^{\rm N} = -3.35$.
Starting from arbitrary SLD values, the reconstruction concludes to values corresponding to real materials:
Mode~1 corresponds to the assumption that isotopic titanium is used for the deposition of the outer layer. In this case, the film surface consists of a wide transition layer Air --$^{48}$Ti ($\eta\approx~120$~\AA ). The distance from the film surface to the Tb$_{0.03}$Co$_{0.97}$ layer is about 400~\AA.
Mode~2, on the other hand, suggests that the outer layer is titanium oxide, formed by the oxidation of titanium in the surrounding atmosphere. This titanium oxide layer then passes into a thin layer of pure natural titanium and a thin transition layer ($\eta \approx$ 100 \AA) before reaching the Tb$_{0.03}$Co$_{0.97}$ layer. The combined thickness of these interface layers is about 300 \AA. Note that the obtained values of structural and magnetic parameters for the main magnetic layer Tb$_{0.03}$Co$_{0.97}$ do not vary significantly for the different modes (see Table~1).
Thus, the stochastic approach extends the reconstruction beyond the limits of a single best-fit solution. This makes it possible to find unexpected properties of the structure. This is particularly useful in the case of statistically ambiguous experimental data.

Having the MCMC samples, we calculated Pearson correlation martrices $\mathbf{R}$ for both solutions
(see Fig.~\ref{fig:pearson}).
It is symmetric, with diagonal elements strictly equal to one. 
Thus, only half of the matrix contains unique information. In the Figure \ref{fig:pearson}, 
the lower left corner corresponds to Mode~1 parameters, while the upper right corner corresponds to Mode~2 parameters.
Two modes show different correlation patterns among the parameters, as can be seen in Figure~\ref{fig:pearson}
as well as in the off-diagonal panels in Fig~\ref{fig:corner_plot}. What they have in common is that for both modes the parameters for the interface layers are correlated with each other, while the parameters for the inner magnetic layer are significantly correlated with each other.  One difference is the stronger correlation between the interface and inner magnetic layers, which is characteristic of Mode~1, while such a correlation is not evident for Mode~2.
The pattern of correlations is similar for the parameters describing the magnetic layer in both Mode~1 and Mode~2 solutions. However, the parameters describing the interface are different between Mode~1 and Mode~2. As a visual aid, we have divided these sets of correlations into blocks with dashed lines in Fig.~\ref{fig:pearson}. The substrate SLD and error model parameters have low correlation with all other parameters and with each other. This indicates that both Mode 1 and Mode 2 solutions are insensitive to variations in these parameters. Note also that the magnetic SLD of the TbCo layer is strongly negatively correlated in both modes (Mode~1 $R = -0.89$ and Mode~2 $R = -0.95$). Of course, this is because the model preserves the total amount of scattering capability of the given layer, i.e. the reduction of the magnetic SLD can be compensated by the nuclear SLD. This is a source of ambiguity in the reconstruction. However, it is still possible to distinguish between magnetic and nuclear SLDs, since they correlate differently with the thickness of the magnetic layer. In the end, the uncertainty in the reconstruction of the magnetic SLD is less than 2\% (see Table~1). Overall, the reconstruction differs only in the interpretation of the surface structure. This can also be seen in the SLD profiles in Figure~\ref{fig:profs}

Let us discuss the reasons that lead to such ambiguity in the reconstruction of the structure of the studied thin film Tb$_{0.03}$Co$_{0.97}$. An obvious reason is the limited statistical power of the experiment caused by the low neutron flux on the reflectometer. As a result, data points corresponding to wave vector transfer values $Q>0.03$~\AA$^{-1}$ have large uncertainties and contribute minimally to the likelihood function. This, together with the background contribution, makes it difficult to reconstruct the profile unambiguously. Another reason is the inability to determine the exact experimental resolution. As mentioned above, the parameter $\delta Q/Q$ was used in the optimization process due to the impossibility to determine the experimental resolution exactly. The two resolution parameters obtained for Mode~1 and Mode~2, $8.3^{+0.2}_{-0.2}$ and $9.1^{+0.2}_{-0.2}$ respectively, as shown in Figure~\ref{fig:corner_plot}, are consistent with the estimates based on the setup parameters, which predict a resolution of about 8-9\%. A final reason is related to the specificity of neutron scattering - isotope sensitivity. As a result, even significantly different SLD-profiles, such as those shown in Figure~\ref{fig:profs}, can describe the same structure made of different titanium isotopes.

Thus, the stochastic analysis of the PNR data allowed us to study the magnetic properties of the TbCo thin film structure. Moreover, it revealed two possible profiles of the surface structure of the sample. These solutions are plausible with similar probabilities. Coincidentally, two solutions can be attributed to the different isotopic compositions of the surface. In our particular case, we were convinced that the Mode 2 solution was correct because the isotopic composition was controlled in the growth process and the depth of the Air-$^{48}$Ti transition layer for Mode 1 was measured to be 120\AA, which is physically unlikely. However, the main point here is that the MCMC approach is indeed capable of finding multiple structure reconstructions from PNR data. This extends the applicability of PNR to problems in physics and nanometrology.
\section{Conclusion}

In this work, an approach based on Markov chain Monte Carlo sampling and Bayesian statistics was used to analyze statistically ambiguous polarized neutron reflectometry data obtained from thin film Tb$_{0.03}$Co$_{0.97}$. Brief descriptions of the experimental conditions and sample synthesis are provided, along with detailed information on the methodology used and the optimization process.

The presence of multimodality in the solutions has been established, with parameter sets corresponding to significantly different reconstructed profiles exhibiting comparable likelihood function values. Intriguing, both reconstructed profiles were found to be realistic and could potentially correspond to the true structure of the investigated thin film. Pairwise correlation dependencies were analyzed and presented for both sets of solutions, with the most significant correlations observed for structural parameters associated with interfacial layers.
Possible sources of ambiguity in the reconstruction of the film structure are discussed and analyzed.

Thus, the application of Markov chain Monte Carlo sampling and Bayesian statistics to the analysis of polarised neutron reflectometry data has demonstrated its exceptional effectiveness. It is of interest to further apply this method to the analysis of data from neutron reflectometry experiments and to apply this approach to the analysis of data from other neutron experiments, such as small-angle neutron scattering and neutron spectroscopy.

\subsection*{Acknowledgements}
The authors are deeply grateful to  Dr.~N.K.~Pleshanov  and Dr.~V.G.~Syromyatnikov for stimulating interest in the present study and useful discussions of the results. PSS is grateful to S.~Derevyashkin for fruitful discussions on optimization implementation issues.

\bibliographystyle{ieeetr}	
\bibliography{mybib.bib}

\end{document}